# Charge state tuning of spin defects in hexagonal boron nitride

J. Fraunié[1], T. Clua-Provost[2], S. Roux[1], Z. Mu[2], A. Delpoux[1], G. Seine[3], D. Lagarde[1], K. Watanabe[4], T. Taniguchi[5], X. Marie[1], T. Poirier[6], J.H. Edgar[6], J. Grisolia[1], B. Lassagne[1], A. Claverie[3], V. Jacques[2†], C. Robert[1†]

[1]*Université de Toulouse, INSA-CNRS-UPS, LPCNO, 135 Av. Rangueil, 31077 Toulouse, France*
[2]*Laboratoire Charles Coulomb, Université de Montpellier and CNRS, 34095 Montpellier, France*
[3]*CEMES-CNRS and Université de Toulouse, 29 rue J. Marvig, 31055 Toulouse, France*
[4]*Research Center for Electronic and Optical Materials, National Institute for Materials Science, 1-1 Namiki, Tsukuba 305-0044, Japan*
[5]*Research Center for Materials Nanoarchitectonics, National Institute for Materials Science, 1-1 Namiki, Tsukuba 305-0044, Japan*
[6]*Tim Taylor Department of Chemical Engineering, Kansas State University, Kansas 66506, USA*

*Boron vacancies in hexagonal boron nitride (hBN) are among the most extensively studied optically active spin defects in van der Waals crystals, due to their promising potential to develop two-dimensional (2D) quantum sensors. In this letter, we demonstrate the tunability of the charge state of boron vacancies in ultrathin hBN layers, revealing a transition from the optically active singly negatively charged state to the optically inactive doubly negatively charged state when sandwiched between graphene electrodes. Notably, there is a photoluminescence quenching of a few percent upon the application of a bias voltage between the electrodes. Our findings emphasize the critical importance of considering the charge state of optically active defects in 2D materials, while also showing that the negatively charged boron vacancy remains robust against external perpendicular electric fields. This stability makes it a promising candidate for integration into various van der Waals heterostructures.*

*Keywords: color centers, 2D materials, quantum sensing*

Quantum sensing based on optically active spin defects in two-dimensional (2D) materials has garnered significant attention in recent years, driven by the potential to bring sensors into atomic-scale proximity with the samples being probed. Such proximity, which is particularly valuable for detecting weak magnetic field sources, cannot be reached with spin-based quantum sensors hosted in 3D materials, such as nitrogen-vacancy (NV) centers in diamond. Moreover, 2D sensing layers offer integrated tools to investigate the emerging physics of complex van der Waals (vdW) heterostructures. Among various 2D materials, color centers in hexagonal boron nitride (hBN) are the most extensively studied, owing to the host material's large band gap (~6 eV) and its successful integration into van der Waals heterostructures[1,2]. While numerous color centers emitting light across the ultraviolet to near-infrared spectrum have been identified in hBN[3], only a select few exhibit the desirable spin properties necessary for quantum sensing[4–9].

The negatively-charged boron vacancy ($V_B^-$) center is the most prominent among these. Under green laser excitation, it emits a photoluminescence (PL) signal centered at 850 nm, which is perfectly photostable, even under ambient conditions. Moreover, the $V_B^-$ center in hBN exhibits a spin triplet ground state whose electron spin resonance (ESR) frequencies can be interrogated by optical means[10]. In recent years, this property has been exploited for sensing magnetic fields, strain, or temperature[10]. However, most current proofs of concept of quantum sensing with $V_B^-$ centers have relied on relatively thick hBN layers (on the order of tens of nanometers)[11–14]. In a previous work, we showed that while the PL properties of $V_B^-$ are preserved in ultra-thin layers[15], the zero-field splitting of the ESR spectrum disappears, and longitudinal spin relaxation is reduced[16]. Another critical aspect to consider in thin layers is the stabilization of the defect's charge state. The charge state of NV centers, for example, can change when the defect is positioned close to a surface due to band bending effects[17–19]. For the boron vacancy, only the singly negatively charge state ($V_B^-$) displays a spin dependent PL response. Thus, it is essential to investigate how the PL intensity varies from modifications to the charge state in thin hBN layers.

Although significant theoretical work has been done on the charge states of color centers in hBN [20–23], experimental investigations remain limited. The charge state of visible single-photon emitters in hBN can be modulated by external gating [24–26] or by functionalizing adjacent graphene layers[27]. Gong et al. estimated the ratio of negatively charged boron vacancies ($V_B^-$) to the total boron vacancy density as a function of ion implantation fluence[28]. At high fluences, the ratio is as low as 1%, suggesting that PL from $V_B^-$ could be enhanced through external gating. In contrast, Gale et al. demonstrated that negatively charged boron vacancies ($V_B^-$) can be ionized under electron beam irradiation, but an external gating had no effect in the absence of the electron beam[29].

In this letter, we demonstrate the ability to tune the charge state of the $V_B$ defect in thin hBN layers (less than 15 nm) sandwiched between few-layer graphene (FLG) electrodes. Notably, we observe a small PL quenching as a function of bias, which suggests that, at zero bias, most vacancies are singly negatively charged. By fabricating structures with an inhomogeneous distribution of vacancies, we further show that $V_B^-$ centers near the FLG electrodes in the hBN layers transition to the $V_B^{2-}$ state under the application of bias.

The first structure studied is illustrated in Figure 1a, with an optical image presented in Figure 1b. It consists of a 15 nm thick hBN flake sandwiched between two FLG flakes. The FLG flakes were mechanically exfoliated from a highly ordered pyrolytic graphite (HOPG) crystal and transferred onto a pre-patterned Si/SiO$_2$ (80 nm) surface using PDMS stamping[30]. The hBN flake was exfoliated from a bulk crystal synthesized via metal flux growth methods[31], isotopically purified with $^{10}$B and $^{15}$N and subsequently irradiated with neutrons with a fluence of 1.4x10$^{17}$ neutrons.cm$^{-2}$ to create a homogeneous distribution of boron vacancies at approximately 100 ppm[32]. Continuous-wave PL was measured at room temperature, using a 532 nm laser beam focused with a high numerical aperture objective (NA = 0.82). Luminescence from the $V_B^-$ defects was collected through the same objective and detected using a single-photon avalanche diode (SPAD). The laser excitation power was set at 1 mW, resulting in a maximum detection count of around 150 kcounts/s for this sample. We select the signal originating exclusively from $V_B^-$ defects by using a combination of longpass (725 nm) and shortpass filters (900 nm). This is proved by the characteristic ESR spectrum shown in the Supporting Information. Throughout our study, we grounded the top FLG electrode. The PL intensity as a function of the applied bias is shown in Figure 1c (orange points). The luminescence peaked at 0 V and decreased symmetrically for both negative and positive biases, with a maximum quenching of 3% at ±3 V. Importantly, this PL quenching was only detected in the hBN region sandwiched between the two FLG electrodes (see Supplementary Information). The bias was limited to ±3 V, corresponding to an electric field of 0.2 V/nm, which is below the breakdown electric field reported for pristine hBN (~0.5 V/nm)[33]. At higher electric fields, we observed sudden and random degradation of the structure. We confirmed these findings with a second sample featuring a thinner hBN flake (9.5 nm), with results depicted in Figure 1c (red points). To facilitate direct comparison between the two samples, we normalized the PL intensity to its maximum, as the total PL intensity was affected by the different thicknesses of the FLG electrodes. For the same bias, quenching is more pronounced in thinner flakes. Finally, we performed similar measurements on a third device with a comparable thickness (10 nm) but a lower defect density (with a neutron irradiation fluence decreased by one order of magnitude). The results are presented as open circles in Figure 1c and demonstrate that the quenching percentage depends solely on the hBN thickness, rather than the defect density.

Since the results in Figure 1c are consistently reproduced across multiple bias sweeps, we conclude that the PL quenching is not due to the migration of defects within the hBN. Additionally, we rule out the opening of new non-radiative de-excitation channels when a bias is applied, as indicated by the measured decay times of the excited state. Figure 1d presents the time-resolved photoluminescence (TRPL) dynamics for the first sample (15 nm thick, $1.4 \times 10^{17}$ neutrons.cm$^{-2}$ of fluence) at three applied biases (0 V and ±3 V). These experiments were conducted at room temperature using a ps pulsed laser at 540 nm with a repetition rate of 80 MHz and an average power of 500 µW. Detection was performed with a streak camera in synchroscan mode. All three decay curves are identical, exhibiting a decay time of approximately 500 ps, consistent with previous measurements on non-gated flakes[34]. The quenching in the integrated PL intensity with bias, coupled with unchanged decay times, suggests the absence of new non-radiative channels and indicates a reduction in the number of optically active defects due to changes in their charge states.

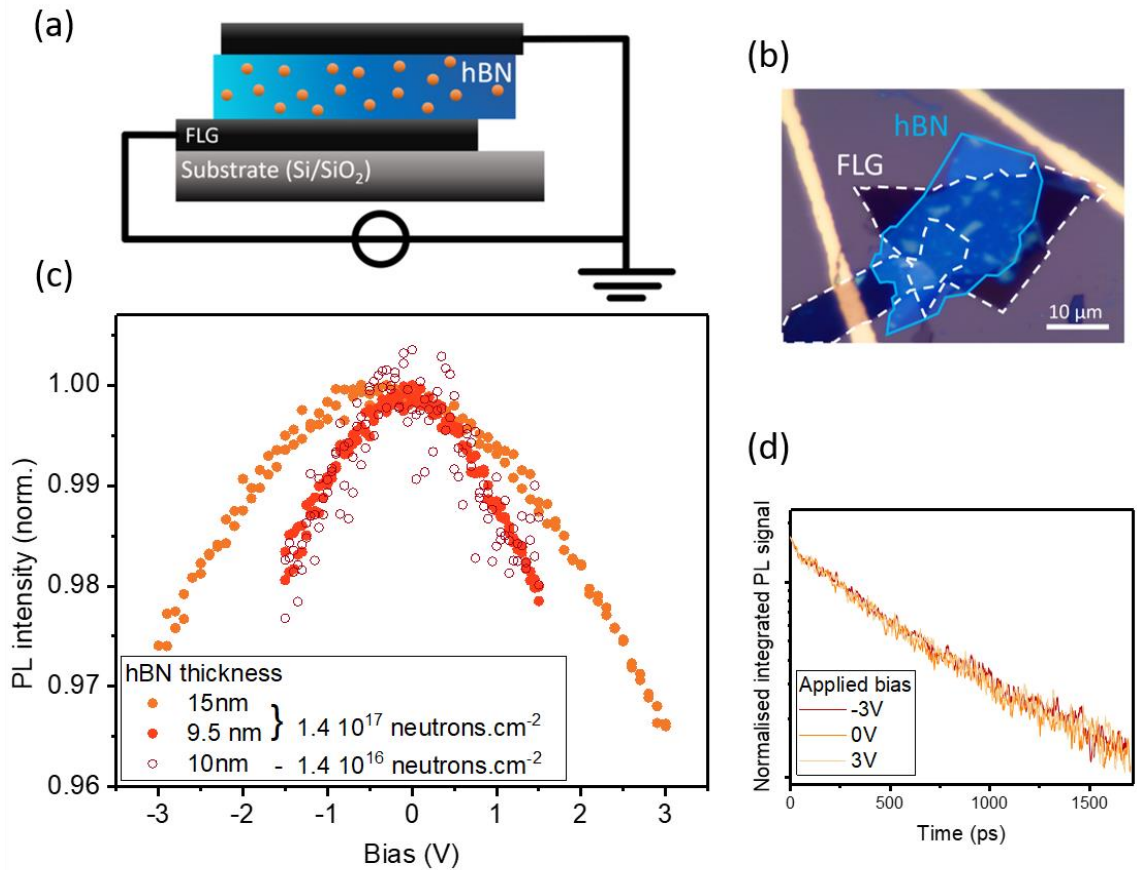

*Figure 1: (a) Sketch of the structure. The hBN flake exfoliated from a bulk irradiated with neutrons is sandwiched between two few layer graphene (FLG) flakes. The defect density is assumed to be homogeneous in this structure. (b) Optical image of the sample. (c) Normalized PL intensity as function of bias of three samples with various hBN thickness and irradiation time. (d) Time-resolved PL decay recorded for the sample 15 nm thick, $1.4 \times 10^{17}$ neutrons.cm$^{-2}$ irradiation fluence as a function of the applied bias.*

At first glance, our findings appear to contradict the results of Gale *et al.*, who reported no change in PL intensity with applied gate voltage[29]. However, our study focuses on much thinner flakes (<15 nm), while Gale *et al.* worked with an 80 nm thick flake. Importantly, we carefully checked that the application of the bias does not lead to any leakage current through the hBN layer (see Supplementary Information). We interpret the modest PL quenching of just a few percent observed in Figure 1c, along with the lack of leakage current under applied bias, as a result of changes in the charge states of boron vacancies located near the FLG electrodes. In this context, charge carriers can only hop between the electrodes and the nearby $V_B$ centers. Importantly, the degree of quenching for a given bias depends primarily on the surface-to-volume ratio of the flake (which is inversely proportional to its thickness), rather than on the defect density. A more detailed discussion of this is provided in the Supplementary Information. Consequently, it is not surprising that quenching is not observed in thicker layers[29], where the ratio of optically active defects ($V_B^-$) to optically inactive defects ($V_B$ in different charge states) remains relatively high.

We now determine the charge state of boron vacancies in hBN. Theoretical studies predict the charge transition levels (CTL) for these centers, revealing that they can exist in three stable charge states[20,21]: neutral ($V_B^0$), singly negatively charged ($V_B^-$), and doubly negatively charged ($V_B^{2-}$), depending on the position of the Fermi level within the hBN band gap. To investigate whether the PL quenching seen in Figure 1c results from a transition from $V_B^-$ to $V_B^0$ or from $V_B^-$ to $V_B^{2-}$, we fabricated two structures with an asymmetric distribution of $V_B$ centers along the flake thickness. The first structure consists of a hBN flake in which boron vacancies were generated through collisions with an ionic beam. A pristine 18 nm hBN flake, exfoliated from a bulk crystal grown under high pressure and temperature conditions[35], was transferred onto a 10 nm thick FLG back electrode deposited on a $Si/SiO_2$ (80 nm) substrate. The structure was then implanted with nitrogen ions at an energy of 30 keV, with a dose of $10^{14}$ ions/cm² to create $V_B^-$ centers[36] The slowing down of the ions and the resulting generation of damage in the structure was simulated by Monte Carlo using the SRIM-2013 code under "full cascade" conditions[37]. Due to the thinness of the hBN flake and the relatively high implantation energy, most nitrogen atoms stop in the $SiO_2$ layer, while boron vacancies are formed in the hBN layer through rare collisions of the incoming ion beam and mostly from recoil atoms. Figure 2a shows the density of vacancies within the overall structure while Figure 2b focuses on the density of boron vacancies in the hBN layer. This simulation shows that the number of vacancies on the bottom side of the hBN flake is roughly twice that found on the top side. After ion implantation, a thin top FLG electrode is deposited. Figure 2c displays the PL intensity of $V_B^-$ as a function of bias with the top electrode grounded. In contrast to the results in Figure 1c, the PL quenching shows asymmetric behavior, with significantly stronger quenching under negative bias. Additionally, quenching with a positive bias begins at around +2.5 V.

To confirm that the asymmetry observed in Figure 2c is linked to the inhomogeneous distribution of $V_B$ centers along the flake thickness, we fabricated a second structure in which boron vacancies are localized near the top electrode. Specifically, we stacked a thin neutron-irradiated hBN flake (4 nm) on top of a pristine hBN flake (15 nm thick), with this stack sandwiched between FLG electrodes. Figure 2d shows the PL intensity as a function of bias. Here, the observed asymmetry is opposite to that in Figure 2c, with PL quenching occurring only under positive bias. Notably, no quenching is detected for negative bias, which aligns with the absence of boron vacancies on the back-electrode side. This contrasts with the structure in Figure 2c, where a few vacancies were present on the top-electrode side.

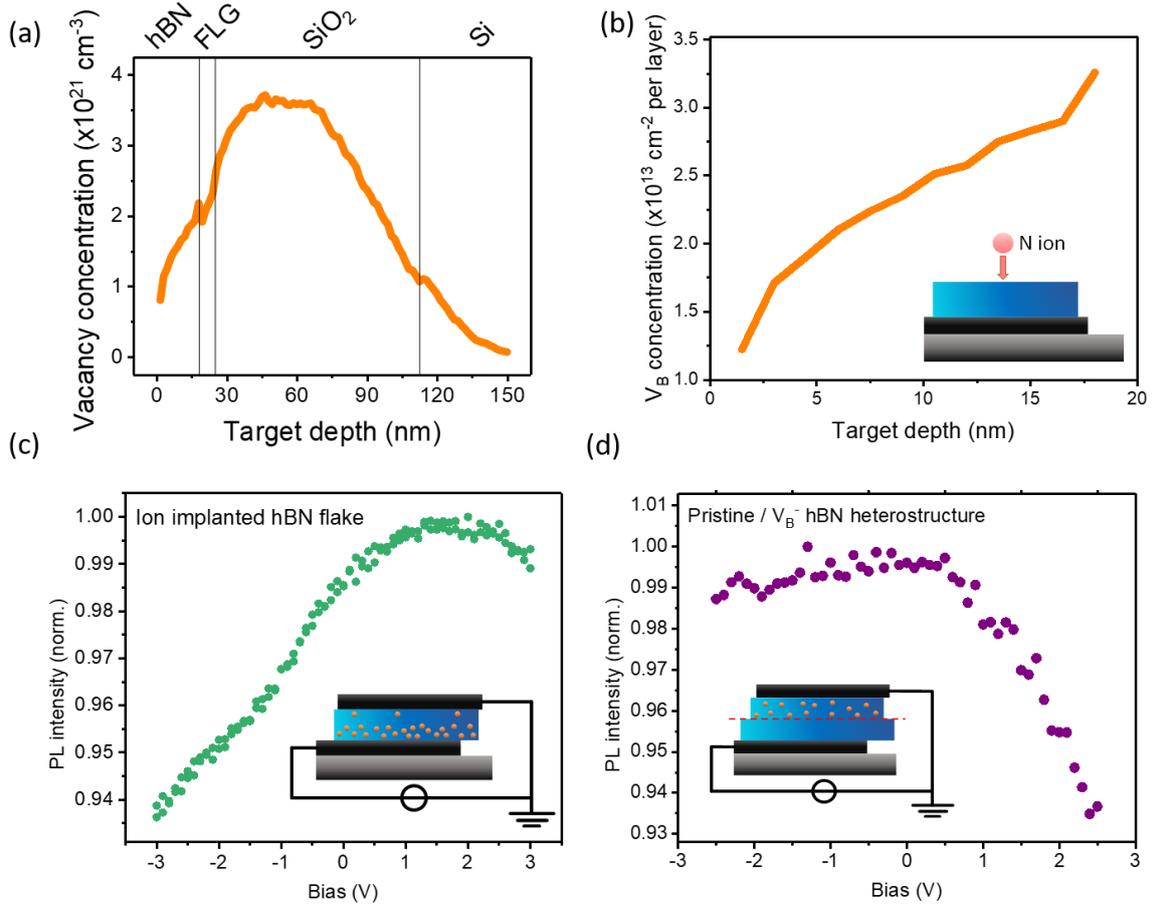

*Figure 2: (a) Monte-Carlo simulation of the number of vacancies (C, B, N, Si and O) generated by an N ion beam in the $hBN_{18nm}/graphite_{10nm}/SiO2_{80nm}/Si$ structure. (b) Focus on the number of B vacancies in the hBN flake. (c) Normalized PL intensity as function of bias of the ion implanted sample. The insert shows the sketch of the structure with an inhomogeneous distribution of defects. (d) Normalized PL intensity as function of bias of a neutron irradiated flake transferred on top of a pristine flake.*

We tentatively interpret our results using the simplified band energy diagram shown in Figure 3, considering the FLG/hBN/FLG stack as a metal-insulator-metal structure. Since we are dealing with vdW interfaces that have no dangling bonds, we can neglect the Fermi level pinning effect typically seen in conventional 3D materials[38]. Thus, the band diagram can be constructed using the simple Schottky-Mott rule. Key parameters in this construction include the workfunctions of hBN and HOPG, which are used for the FLG electrodes, as well as the electron affinity ($E_A$) and band gap ($E_g$) of hBN, and the position of the charge transition levels (CTL). The workfunction of HOPG is well-documented, with a value of $\varphi_G = 4.7 \pm 0.1$ eV[39]. The workfunction of pristine hBN is reported to be $\varphi_{BN}^P = 4.9 \pm 0.1$ eV[40]. However, we expect $\varphi_{BN}$ to vary from its pristine value in the presence of point defects. To quantify this variation, we performed Kelvin Force Microscopy (KFM) on several hBN flakes: pristine hBN, hBN irradiated with neutrons, and hBN implanted with nitrogen ions. All flakes were deposited on the same SiO2/Si substrate and measured under identical conditions with the same AFM tip, allowing us to directly compare their workfunctions. The results, shown in Figure 3a, reveal that the workfunction of hBN with defects ($\varphi_{BN}^D$) — for both irradiated and implanted samples — is ~0.2 eV smaller than that of pristine hBN, i.e., $\varphi_{BN}^D = \varphi_{BN}^P - 0.2$ eV. Based on this, we assume that $\varphi_{BN}^D \cong \varphi_G$ and neglect any band bending at the interface. The experimental band gap of hBN is typically reported as 5.95 eV[41] at low temperature, which corresponds to the optical band gap (energy of the exciton). The free carrier band gap ($E_g$) is generally larger by an amount equal to the binding energy of the exciton, estimated to be around 300 meV[42,43]. Assuming a reduction of 50 meV in the band gap from low to room temperature, we choose $E_g$ = 6.2 eV. Data on the electron affinity of hBN is less abundant and can vary depending on crystal quality[44]. Reported values range from positive affinities of 1.7 eV to 2.3 eV[44–46], with some reports suggesting a negative affinity of -0.5 eV[47]. The position of the CTL has been calculated by Strand *et al.*[21] and Weston *et al.*[20]. For the 0/-1 charge state, it ranges from 1.48 eV to 2.1 eV, and for the -1/-2 charge state, it spans 4.9 eV to 5.2 eV relative to the valence band maximum (VBM). Given these uncertainties in both the electron affinity and the CTL position, it is challenging to construct a precise quantitative band diagram. However, what remains important is the relative position of the Fermi level to the CTL. Two extreme scenarios are illustrated in Figure 3b and 3c, corresponding to two different parameter sets. In both cases, the Fermi level lies between the two CTLs, consistent with a stable singly negatively charge state ($V_B^-$) at 0 V. In Figure 3b, the Fermi level is closer to the CTL 0/-1, while in Figure 3c, it is closer to the CTL -1/-2. Our experimental data for asymmetric structures, shown in Figure 2, help us select the more likely scenario. Figure 3d presents the band diagram for a positive bias. Under this condition, electrons from the top gate electrode can tunnel to the nearest $V_B^-$ defects, changing their charge state to $V_B^{2-}$, while holes from the bottom gate electrode can tunnel to the nearest $V_B^-$ defects, changing their charge state to $V_B^0$. Experimentally, we observed that when defects are situated near the top (bottom) electrode, the PL is quenched for positive (negative) bias (Figure 2d and Figure 2c respectively). This observation supports the idea that the defects transition more readily from $V_B^-$ to $V_B^{2-}$.

We now address the shift in the PL maximum observed at +1.5 V in the ion-implanted sample (Figure 2c), compared to the maximum PL at 0 V in the neutron-irradiated sample. We propose that this shift is due to the asymmetric band bending between the two FLG/hBN interfaces in the ion-implanted sample. Specifically, as shown in the collision simulation in Figure 2b, the bottom side of the hBN flake is more heavily doped than the top side. As a result, the workfunction of the implanted hBN is smaller on the bottom surface than on the top surface. Note that this difference cannot be directly measured using KFM, as it only provides an averaged measurement of the workfunction across the entire flake. Consequently, our previous assumption of negligible band bending at both interfaces is less applicable at the bottom interface. A positive bias is therefore required to counteract this asymmetry and restore the flat-band configuration shown in Figure 3c, leading to the maximum PL intensity.

In conclusion, we have demonstrated that the PL intensity associated with $V_B^-$ defects in very thin layers of hBN can be effectively modulated by applying a bias across external FLG electrodes. We show that the PL quenching is inversely proportional to the flake thickness, though it remains modest (a few percent) even under large electric fields. This suggests that thin $V_B^-$ sensing layers can be integrated into vdW heterostructures with applied perpendicular electric fields without significantly altering the optical properties of $V_B^-$ defects. In neutron-irradiated samples, PL is maximized at zero bias, suggesting that the boron vacancy remains stable in its optically active, singly negatively charged state. At first glance, these results might seem to contradict the findings of Gong *et al.*[28], who estimated that the ratio of $V_B^-$ to the total boron vacancies ranges from 1% to 10%, depending on the implantation fluence. However, it is important to note that Gong *et al.*'s density estimation was based on simulations that did not account for clustering effects. We propose that, at high implantation fluence, most boron vacancies aggregate into clusters, with the majority of single boron vacancies remaining negatively charged. Finally, by using asymmetric structures, we demonstrate that the PL quenching observed under finite bias arises from a transition to the doubly negatively charge state ($V_B^{2-}$). Our findings underscore the importance of considering the charge state of spin defects in quantum sensing applications, particularly when these defects are in close proximity to the material being probed.

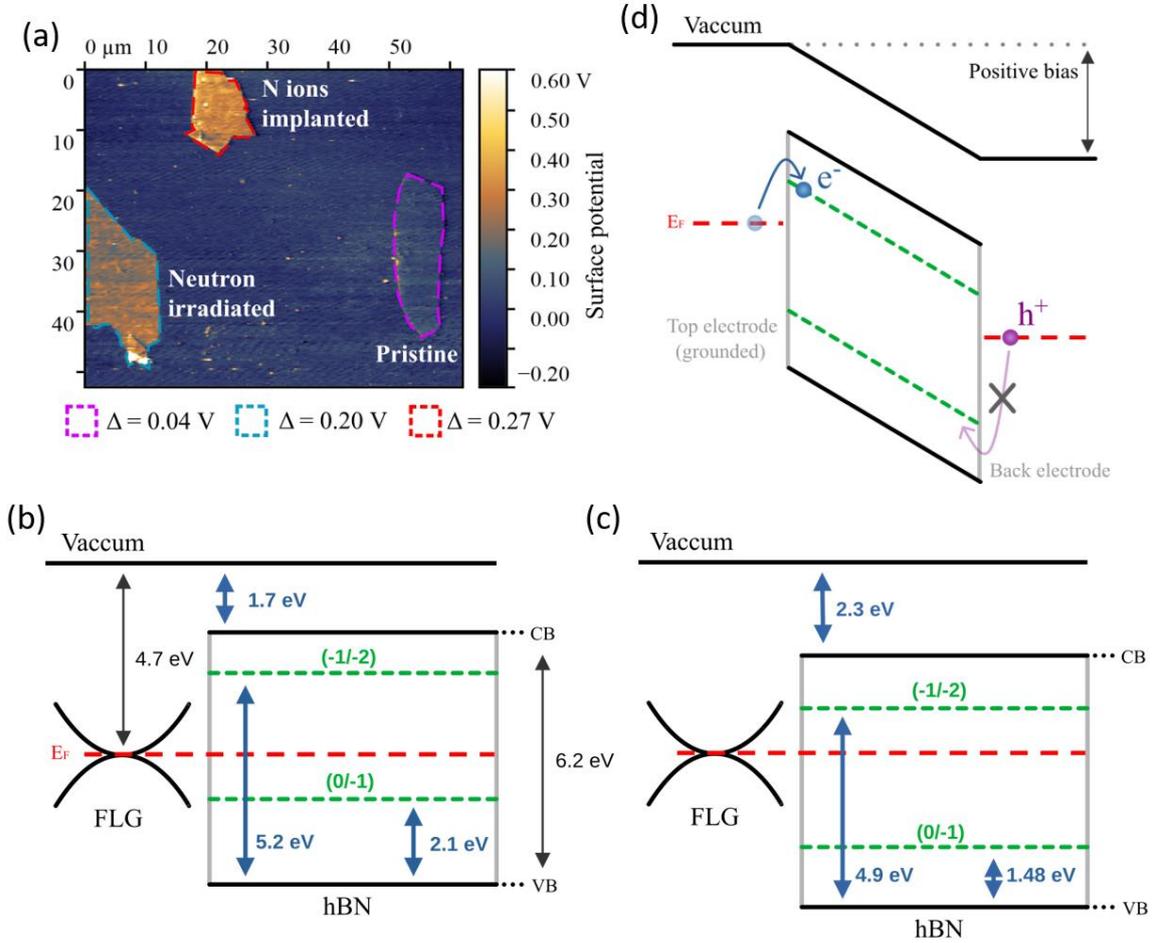

*Figure 3: (a) KFM image of three hBN flakes (pristine, neutron irradiated and ion implanted) showing the variation of their workfunction. (b-c) Simple energy band diagrams between FLG and hBN for two sets of extreme parameters. In (b), the Fermi level is closer to the (0/-1) CTL while it is closer to the (-1/-2) level in (c). (d) Simple energy band diagram of a FLG/hBN/FLG stack under a positive bias. The electrons are more likely to tunnel from the top electrode than the holes from the back electrode.*

**Acknowledgements**

We thank P. Seneor, I. Gerber, C. Zu and N. Yao for fruitful discussions. This work was supported by Agence Nationale de la Recherche funding under the program ESR/EquipEx+ (grant number ANR-21-ESRE-0025), ANR QFoil, the grant NanoX n° ANR-17-EURE-0009 in the framework of the « Programme des Investissements d'Avenir" and the Institute for Quantum Technologies in Occitanie through the project 2D-QSens. K.W. and T.T. acknowledge support from the JSPS KAKENHI (Grant Numbers 21H05233 and 23H02052), the CREST (JPMJCR24A5), JST and World Premier International Research Center Initiative (WPI), MEXT, Japan. JHE and TP acknowledge the support from the Office of Naval Research award number N00014-22-1-2582 for hBN crystal growth. Neutron irradiation of the hBN crystals was supported by the U.S. Department of Energy, Office of Nuclear Energy under DOE Idaho Operations Office Contract DE-AC07-051D13417 as part of a Nuclear Science User Facilities experiment.

† Corresponding author: cerobert@insa-toulouse.fr, vincent.jacques@umontpellier.fr